\documentclass[sigconf]{acmart}

\AtBeginDocument{%
  \providecommand\BibTeX{{%
    \normalfont B\kern-0.5em{\scshape i\kern-0.25em b}\kern-0.8em\TeX}}}

\setcopyright{acmcopyright}
\copyrightyear{2018}
\acmYear{2018}
\acmDOI{XXXXXXX.XXXXXXX}

\acmConference[Conference acronym 'XX]{Make sure to enter the correct
  conference title from your rights confirmation emai}{June 03--05,
  2018}{Woodstock, NY}
\acmPrice{15.00}
\acmISBN{978-1-4503-XXXX-X/18/06}

\usepackage{algorithm}
\usepackage{algpseudocode}
\usepackage{multirow}

\newcommand\modified[1]{\textcolor{blue}{#1}}
\renewcommand\modified[1]{{#1}}

\begin{document}

\title{DGEMM on Integer Matrix Multiplication Unit}


\author{Hiroyuki Ootomo}
\email{ootomo.h@rio.gsic.titech.ac.jp}
\orcid{1234-5678-9012}
\affiliation{%
  \institution{Tokyo Institute of Technology}
  \state{Tokyo}
  \country{Japan}
}
\author{Katsuhisa Ozaki}
\email{ozaki@sic.shibaura-it.ac.jp}
\orcid{1234-5678-9012}
\affiliation{%
  \institution{Shibaura Institute of Technology}
  \state{Saitama}
  \country{Japan}
}
\author{Rio Yokota}
\email{rioyokota@gsic.titech.ac.jp}
\orcid{1234-5678-9012}
\affiliation{%
  \institution{Tokyo Institute of Technology}
  \state{Tokyo}
  \country{Japan}
}

\renewcommand{\shortauthors}{Ootomo and Ozaki, et al.}

\begin{abstract}
\modified{
Deep learning hardware achieves high throughput and low power consumption by reducing computing precision and specializing in matrix multiplication.
For machine learning inference, fixed-point value computation is commonplace, where the input and output values and the model parameters are quantized.
Thus, many processors are now equipped with fast integer matrix multiplication units (IMMU).
It is of significant interest to find a way to harness these IMMUs to improve the performance of HPC applications while maintaining accuracy.
We focus on the Ozaki scheme, which computes a high-precision matrix multiplication by using lower-precision computing units, and show the advantages and disadvantages of using IMMU.
The experiment using integer Tensor Cores shows that we can compute double-precision matrix multiplication faster than cuBLAS and an existing Ozaki scheme implementation on FP16 Tensor Cores on NVIDIA consumer GPUs.
Furthermore, we demonstrate accelerating a quantum circuit simulation by up to 4.33 while maintaining the FP64 accuracy.
}
\end{abstract}
%




\received{20 February 2007}
\received[revised]{12 March 2009}
\received[accepted]{5 June 2009}

\maketitle

\section{Introduction}
The development of processors for machine learning is progressing rapidly, and researchers are exploring their potential for use in other HPC applications \cite{abdelfattah_survey_2021}.
These processors often use low/mixed-precision floating-point, or low-bit-length integer computing units, such as NVIDIA Tensor Cores \cite{nvidia_nvidia_2022} and Google TPUs \cite{jouppi_-datacenter_2017,jouppi_tpu_2023,jouppi_ten_2021}, to achieve high throughput.
While the training of deep learning models typically requires floating-point types like FP16, Bfloat16, TF32, and FP32, the inference involves quantizing model parameters and input/output values to use lower-bit-length integer formats.
This reduces data size, processor circuit area, and power consumption \cite{horowitz_11_2014,jouppi_ten_2021}.
As a result, high-performance processors like NVIDIA GPUs and edge devices like Google Coral Edge TPUs \cite{seshadri_evaluation_2022} are equipped with low-bit-length integer matrix multiplication units (IMMU).
For instance, NVIDIA GPUs provide a {\tt DP4A} instruction that can efficiently compute the inner product of two length-4 8-bit integer (INT8) vectors and accumulate the result in a 32-bit integer (INT32).
In addition, NVIDIA Tensor Cores support the multiplication of INT8 matrices with INT32 accumulation from the Turing architecture and the multiplication of 4-bit integer ({\tt INT4}) matrices from the Ampere architecture.
These integer Tensor Cores can achieve a theoretical peak performance that is $2 \sim 4$ times faster than floating-point Tensor Cores.
Other processors, such as Google TPU v1 \cite{jouppi_-datacenter_2017}, Intel AMX-INT8 \cite{intel_corporation_intel_2022}, and Groq TSP \cite{ahmed_answer_2022}, also support the multiplication of INT8 matrices with INT32 accumulation.
In light of these advancements, our research aims to investigate whether these processors can be leveraged for other high-performance computing (HPC) applications.

Recent research has explored the use of machine learning processors for HPC applications, in particular low- and mixed-precision floating-point matrix multiplication units (FMMU).
For instance, Haidar \textit{et al.} used the FP16 Tensor Core to solve FP64 linear equations with an iterative refinement method \cite{haidar_harnessing_2018}.
Finkelstein \textit{et al.} have employed FP16 Tensor Cores to perform time-independent quantum response calculations with sufficient accuracy \cite{finkelstein_quantum_2022}.
Our previous study has utilized FP16 Tensor Cores to improve the throughput of quantum circuit simulation by emulating single-precision matrix multiplication using an error correction method and avoiding rounding inside Tensor Cores without loss of accuracy \cite{ootomo_recovering_2022,ootomo_quantum_2023}.
However, despite the promising results obtained by these studies, few have explored the potential of integer matrix multiplication units for HPC applications.
This paper investigates the feasibility of leveraging integer matrix multiplication units, for instance, INT8 Tensor Cores in NVIDIA GPUs, for HPC applications.


Even before the advent of deep learning and specialized hardware for it, researchers had been exploring ways to perform high-precision matrix multiplication on lower-precision computing units.
One such approach is the double-double and quad-double precision, where a single value is represented as the sum of two and four FP64 values, respectively, and arithmetic operations are performed using a sequence of FP64 operations \cite{hida_algorithms_2001}.
GEMM and other BLAS functions for double-double precision have been evaluated on NVIDIA GPUs and AMD Cypress GPUs \cite{mukunoki_implementation_2012,nakasato_fast_2011}.
Another approach is the Ozaki scheme, which also splits a value into multiple lower-precision values \cite{ozaki_error-free_2012,ozaki_generalization_2013}.
In this scheme, input matrices are split into multiple matrix slices, and the resulting matrix is obtained by accumulating the multiplication of these slices.
The key feature of this scheme is that it splits the matrices in a way that prevents rounding errors during each multiplication of the slices, which can be computed on lower-precision computing units.
The resulting matrix can then be obtained by summing up the result of each multiplication of the slices in high precision.
Mukunoki \textit{et al.} have evaluated the Ozaki scheme on many-core processors, including NVIDIA GPUs \cite{mukunoki_reproducible_2020}, and combined it with FP16 Tensor Cores to compute double-precision matrix multiplication.
They showed that their implementation outperformed the theoretical peak performance of FP64 on NVIDIA TITAN RTX \cite{mukunoki_dgemm_2020}.

This paper explores the possibility of using integer matrix multiplication units for the Ozaki scheme.
Our study reveals that leveraging these units can offer more than just faster throughput compared to floating-point matrix multiplication units.
In particular, we show the theoretical advantages in terms of accuracy, computational complexity, and memory consumption.
To evaluate the throughput, we use NVIDIA consumer GPUs in addition to A100 GPU, which are not designed for HPC applications but have high-throughput integer matrix multiplication units, and show that we can run HPC applications fast even on these processors.


The summary of our contributions is as follows:
\begin{itemize}
    \item We show the theoretical advantages of using the integer matrix multiplication unit instead of floating point one concerning the accuracy, memory consumption, and the number of operations.
    It reduces the $50$\% $\sim 75$\% of working memory, a major concern in the Ozaki scheme in practical use, in the middle $\sim$ large size of matrix multiplication.
    \item We implement the Ozaki scheme on NVIDIA integer Tensor Cores and evaluate the accuracy, throughput, and power efficiency.
    We compare our implementation to cuBLAS DGEMM and an existing implementation on FP16 Tensor Cores by Mukunoki \textit{et al.} \cite{mukunoki_dgemm_2020}.
    \modified{Although there is a tradeoff between the throughput and the width of the exponent distribution of the input matrices, our implementation outperforms cuBLAS DGEMM and the existing implementation up to about $6 \times$ on NVIDIA consumer GPUs.}
    \item We apply the Ozaki scheme on integer Tensor Cores to quantum circuit simulation and evaluate the accuracy and throughput.
    For practical use, we implement an automatic accuracy tuning mechanism.
    \modified{We have achieved up to $4.33 \times$ throughput improvement compared to cuBLAS ZGEMM computation on NVIDIA RTX6000 Ada GPU while maintaining the FP64 accuracy.}
\end{itemize}

\section{Background}
\subsection{Floating point arithmetic and rounding}
\label{sec:fp-arith}
A floating point value, for instance, FP32 (Binary32) and FP64 (Binary64), consists of three parts: a sign bit, an exponent part, and a mantissa part.
Then a normalized value $v$ in a floating point value set is represented as follows:
\begin{equation}
    v = (-1)^{s} \times \underbrace{1.m_1 m_2 \cdots m_{(\ell_m-1)}}_\text{mantissa part (binary number)} \times 2^{e},
\end{equation}
where $s$ is the sign bit and $e$ is the exponent part, a signed integer.
For instance, in the case of FP64, the exponent part bit length is $11$ bit, and the mantissa part bit length is $\ell_m = 53$.
The addition of two floating point values is conducted as follows:
\begin{enumerate}
   \item Right-shift the mantissa of the smaller absolute value to match the digit of the larger one.
   \item Add the two mantissae as integers.
   \item Round the added mantissa to be the length of the output mantissa.
   \item Adjust the output mantissa and exponent to fit the floating point value format.
\end{enumerate}
The error by cutting the mantissa in (3) is referred to as a ``rounding error.''
There are several rounding modes in (3), for instance, Round to Nearest; ties to even (RN), Round toward Zero (RZ), etc.
In most modern processors and programming environments, including CUDA, the RN mode is used by default since the expectation of the rounding error is smaller than the others.
The RZ mode simply truncates the mantissa bits.
On the other hand, in the case of the multiplication of two floating point values, we multiply the mantissae of the inputs as integers and do (3) and (4) as same as addition.

\subsection{High precision matrix multiplication on lower precision computing unit}
\begin{figure}
    \centering
    \includegraphics[width=\linewidth]{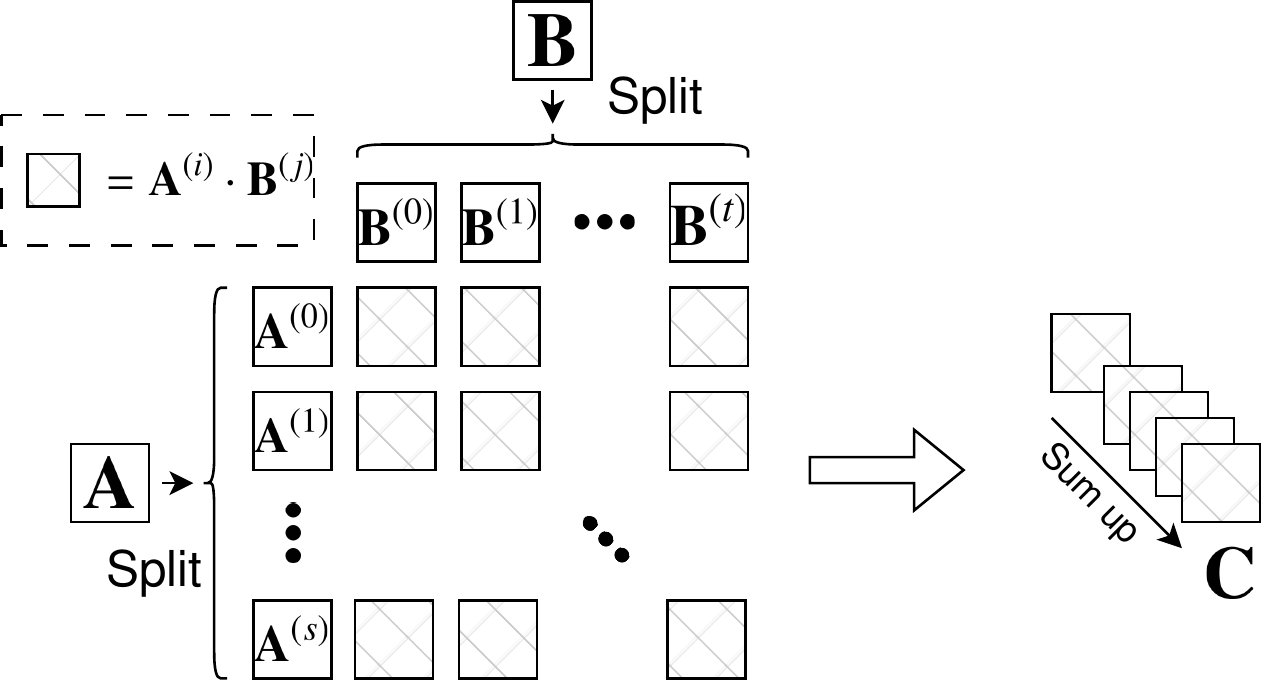}
    \caption{The basic computation of high precision matrix multiplication $\mathbf{C}\leftarrow\mathbf{A}\cdot\mathbf{B}$ algorithms on low precision computing unit.}
    \label{fig:split-gemm-algo}
\end{figure}
There are two kinds of high-precision matrix multiplication methods on lower-precision units.
We refer to them as the elementwise-place splitting method and shared-place splitting method.
Although both methods split the mantissa of the input matrices and sum up the multiplications of these subdivided matrices, as shown in Fig. \ref{fig:split-gemm-algo}, the splitting mechanisms are different.

\subsubsection{elementwise-place splitting method}
The elementwise-place splitting method splits a high-precision value into multiple lower-precision values by only looking at the target value.
The method includes double-double, quad-double \cite{hida_algorithms_2001}, Henry's method using BF16 on Intel's mixed-precision FMA units \cite{henry_leveraging_2019}, and our method using FP16 or TF32 on Tensor Cores \cite{ootomo_recovering_2022}.
Fasi \textit{et al.} conducted the error analysis of the elementwise-place splitting method using two FP16 to emulate the single precision accuracy \cite{fasi_matrix_2023}.
The basic computation is as follows.
We consider an inner product of two vectors instead of matrix multiplication since matrix multiplication can be regarded as a set of inner products.
When computing an inner product of two high-precision length-$k$ vectors $\mathbf{a}$ and $\mathbf{b}$, we split them into $\mathbf{a}^{(1)}, \mathbf{a}^{(2)}, \cdots ,\mathbf{a}^{(s)}$ and $\mathbf{b}^{(1)}, \mathbf{b}^{(2)}, \cdots ,\mathbf{b}^{(t)}$, where
\begin{align*}
    \mathbf{x}^{(1)} = \mathrm{toLower}\left(\mathbf{x}\right), \mathbf{x}^{(i)} = \mathrm{toLower}\left(\mathbf{x}-\sum_{j=1}^{i-1}x^{(j)}\right),
\end{align*}
and ``toLower'' is a precision conversion to lower.
Then we compute the resulting inner product by summing up the inner products of the subdivided vectors as follows:
\begin{equation}
    \label{eq:accumulation-sub-inner}
    \mathbf{a}^\top \cdot \mathbf{b} \approx \sum_{i=1}^{s} \sum_{j=1}^{t} {\mathbf{a}^{(i)}}^\top \cdot \mathbf{b}^{(j)}.
\end{equation}
Intuitively, this method keeps top $\alpha$ bits of the mantissa of the elements of $\mathbf{x}$ by $\mathbf{x}^{(1)}$, leading $\alpha$ bits by $\mathbf{x}^{(2)}$, ..., where $\alpha$ is the mantissa length of the lower precision.
\modified{To be precise, each element substantially keeps slightly more bits than $\alpha$ when using RN for the rounding.}
Since the rounding error occurs in each ${\mathbf{a}^{(i)}}^\top \cdot \mathbf{b}^{(j)}$ computation, the precision of the accumulation in ${\mathbf{a}^{(i)}}^\top \cdot \mathbf{b}^{(j)}$ limits the resulting accuracy.
For instance, in our method on Tensor Cores \cite{ootomo_recovering_2022}, the accuracy is limited to single-precision since the accumulator of FP16 Tensor Cores they use is FP32.
If there were Tensor Cores that have an FP64 accumulator while the input precision is lower, we could compute the matrix multiplication in double precision by this method.
\modified{And therefore, in contrast to the shared-place splitting method we explain later, we can not improve the accuracy more than the accuracy of the accumulation even if we increase the number of splits to keep more input mantissa.}

\subsection{Ozaki scheme}
\label{sec:ozaki}
\modified{The shared-place splitting method also splits the mantissa of the elements of the input matrices and computes the resulting inner product by Eq. (\ref{eq:accumulation-sub-inner}) as same as the elementwise-place splitting method.
However, the mantissa splitting mechanism is different from the elementwise-place splitting method.
The shared-place splitting method is generalized as the Ozaki scheme.}

\modified{The Ozaki scheme \cite{ozaki_error-free_2012,ozaki_generalization_2013} is a matrix multiplication algorithm using working precision arithmetic that is lower than the matrix multiplication's target precision.
For instance, the Ozaki scheme computes double-precision matrix multiplication using single-precision arithmetic.
}
In this section, we show the intuitive mechanism, algorithm, advantages, and disadvantages of the Ozaki scheme.
\subsubsection{Intuitive mechanism}
As we mention in Sec \ref{sec:fp-arith}, the rounding error occurs in the floating-point arithmetic operations.
The Ozaki scheme splits the mantissa of the input vector so that no rounding error occurs in each ${\mathbf{a}^{(i)}}^\top \cdot \mathbf{b}^{(j)}$ computation in Eq. (\ref{eq:accumulation-sub-inner}).
To compute an inner product of two vectors, we repeat multiplying two values and accumulating the results.
Therefore, we need to avoid the rounding error in both multiplication and accumulation.
In the case of the multiplication of two values, the rounding error does not occur when the valid mantissa length of the input values is short enough, where the valid mantissa length is the length from the most significant bit (MSB) to the last 1 of the mantissa.
In particular, when the resulting mantissa length is $\ell_m$, and the valid mantissa lengths of the inputs are $\hat{\ell}_m^{[1]}$ and $\hat{\ell}_m^{[2]}$, we can multiply the inputs without rounding error if $\hat{\ell}_m^{[1]} + \hat{\ell}_m^{[2]} \leq \ell_m$.
This is derived by the algorithm of multiplication.
In the case of adding two values, the rounding error does not occur when 1) the valid mantissa length of the smaller exponent value is short enough and 2) the difference between the exponents is small enough.
Using these features of the floating-point arithmetics, the Ozaki scheme avoids the rounding error during the computation of ${\mathbf{a}^{(i)}}^\top \cdot \mathbf{b}^{(j)}$.
In the Ozaki scheme, we first split the vector $\mathbf{a}$ and $\mathbf{b}$ into vectors $\mathbf{a}^{(1)}, \cdots, \mathbf{a}^{(s)}$ and $\mathbf{b}^{(1)}, \cdots, \mathbf{b}^{(t)}$, respectively, so that the following conditions are satisfied:
\begin{itemize}
    \item $\mathbf{a} = \sum_{i=1}^{s} \mathbf{a}^{(i)}$ and $\mathbf{b} = \sum_{i=1}^{t} \mathbf{b}^{(i)}$
    \item The rounding error does not occur in the computation of the inner product between each of $\mathbf{a}^{(1)}$, $\cdots$, $\mathbf{a}^{(s-1)}$ and $\mathbf{b}^{(1)},$$ \cdots,$$ \mathbf{b}^{(t-1)}$ even on low-precision computing units.
\end{itemize}
Then, we obtain the final result by summing up the result of each inner product as Eq. (\ref{eq:accumulation-sub-inner}) in the target precision.
\modified{Intuitively, ${\mathbf{a}^{(i)}}^\top \cdot {\mathbf{b}^{(j)}}$ computes the high digit part of the resulting mantissa when $i+j$ is small, for instance, $2$, and low digit part when $i+j$ is large, for instance, $s+t$.}
Note that the rounding error can occur in the summing-up process while not occurring in each inner product computation.

\begin{figure}[t]
    \centering
    \includegraphics[width=\linewidth]{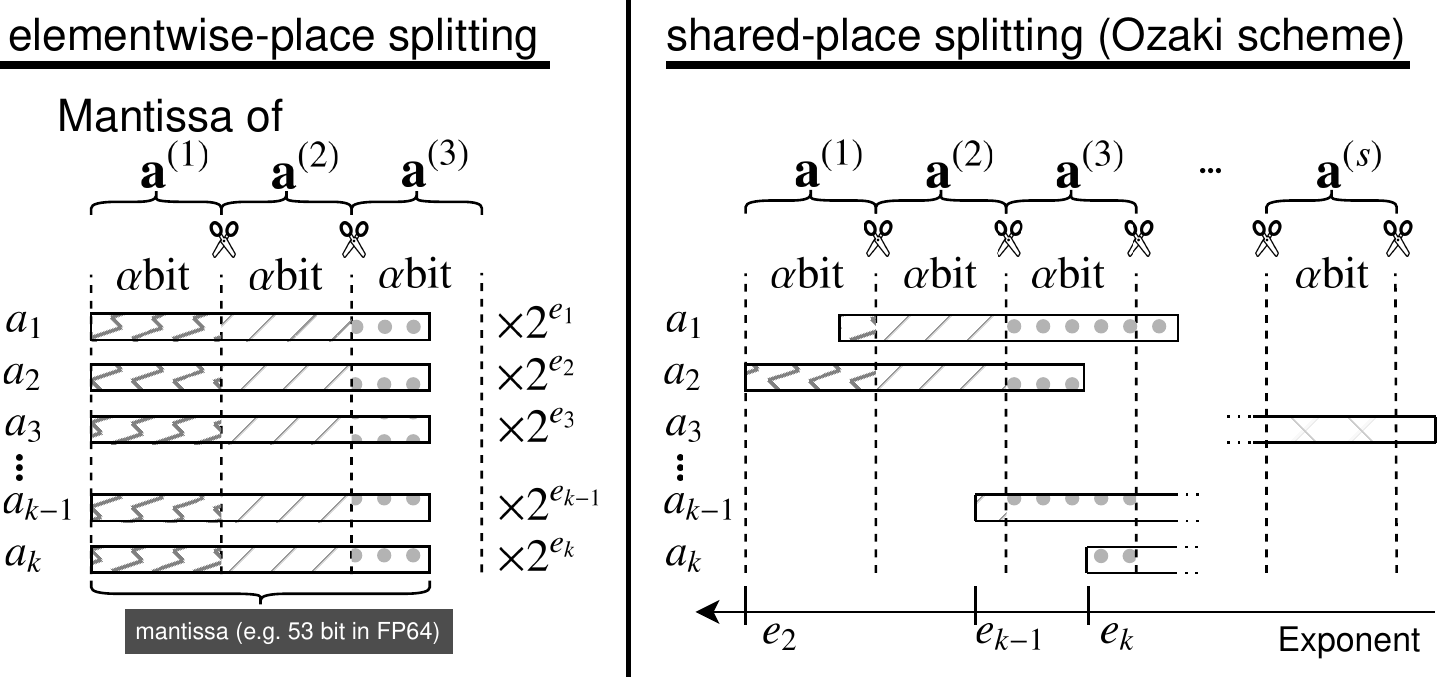}
    \caption{
    The comparison of the elementwise-place splitting (Left) and shared-place splitting (Right) methods for splitting the vector $\mathbf{a}=\begin{bmatrix}a_1 & a_2 & \cdots & a_k\end{bmatrix}^\top$ into several vectors $\mathbf{a}^{(\cdot)}$.
    $e_i$ is the exponent of $a_i$.
    The same applies to the vector $\mathbf{b}$.
    }
    \label{fig:ozaki-scheme}
\end{figure}
In splitting the mantissa, we split the mantissa space of the vector, which is the space of the mantissa, as shown at the right of Fig. \ref{fig:ozaki-scheme}.
The first slice vector ($\mathbf{a}^{(1)}$) keeps the $\alpha$ bits of the space from the most significant mantissa digit in the vector, and the second one ($\mathbf{a^{(2)}}$) keeps the leading $\alpha$ bits.
Then, $\alpha$ is calculated as follows:
\begin{equation}
	\label{eq:alpha}
	\alpha = \lfloor(-\log_2 u_\text{acc} - \log_2 k)/2 \rfloor,
\end{equation}
where $u_\text{acc}$ is the unit round off of the accumulator for computing ${\mathbf{a}^{(i)}}^\top\cdot\mathbf{b}^{(j)}$.
For instance, when computing an inner product of length-$4096$ FP64 vectors by the Ozaki scheme using FP32 ($u_\text{FP32}=2^{-24}$) arithmetic for computing ${\mathbf{a}^{(i)}}^\top\cdot\mathbf{b}^{(j)}$, we split the mantissa spaces by $\alpha=6$ bits.
By splitting the vector in this way, we can satisfy the condition to avoid the rounding error in the operations inside the inner product computation.

\subsubsection{Algorithm}

We show the overall algorithm in Algorithm \ref{alg:ozaki-fp} and the splitting algorithm in Algorithm \ref{alg:ozaki-fp-splitting}.
\modified{We set the number of splits $s=t$ in these algorithms and the later part of this paper.}
When the data types of the input matrix and output matrix are the same, it suffices to sum up $\mathbf{A}^{i}\cdot\mathbf{B}^{(j)}$ for $i + j \leq s + 1$ since the other results only have a negligible effect on the resulting accuracy.

\begin{algorithm}[t]
    \caption{Matrix multiplication by the Ozaki scheme}\label{alg:ozaki-fp}
    \begin{algorithmic}[1]
        \Require Input matrix $\mathbf{A}, \mathbf{B}$, Num split $s$
        \Ensure $\mathbf{C} \leftarrow \mathbf{A}\cdot\mathbf{B}$
        \State $\mathbf{A}^{(1)}, \mathbf{A}^{(2)}, \cdots, \mathbf{A}^{(s)}\leftarrow \text{SplitFP}\left(\mathbf{A}, s\right)$
        \State $\mathbf{B}^{(1)}, \mathbf{B}^{(2)}, \cdots, \mathbf{B}^{(s)}\leftarrow \text{SplitFP}\left(\mathbf{B}, s\right)$
        \State $\mathbf{C} = 0$
        \For $ \text{ \ }i = 1..s$
        \For $ \text{ \ }j = 1..(s-i+1)$
        \State $\mathbf{C}_\text{tmp} \leftarrow \mathbf{A}^{(i)}\cdot\mathbf{B}^{(j)}$ // Low-precision. No rounding error
        \State $\mathbf{C} \leftarrow \mathbf{C} + \mathbf{C}_\text{tmp}$ // High-precision accumulation
        \EndFor
        \EndFor
    \end{algorithmic}
\end{algorithm}

\begin{algorithm}[t]
    \caption{SplitFP: Splitting algorithm of the Ozaki scheme for $m \times k$ input matrix $\mathbf{M}=\mathbf{A} \text{ or } \mathbf{B}^\top$}\label{alg:ozaki-fp-splitting}
    \begin{algorithmic}[1]
        \Require Input $m \times k$ matrix $\mathbf{M}$, Num split $s$
        \Ensure Split matrices list: $\mathbf{M}^{(1)}, \mathbf{M}^{(2)}, \cdots, \mathbf{M}^{(s)}$
        \State $\beta = \lceil(-\log_2 u + \log_2 k) / 2\rceil$
        \State $\mathbf{R} := \mathbf{M}$
        \For \text{ \ }$p = 1..(s-1)$
        \State Compute $\mathbf{m}^{(p)}$ where $\mathbf{m}_i^{(p)} \leftarrow \text{max}_j|\mathbf{R}_{i,j}|$
        \State Compute $\mathbf{\sigma}^{(p)}$ where $\sigma_i^{(p)} \leftarrow 0.75 \times 2^{\lceil \log_2 m_i^{(p)} \rceil} \times 2^\beta$ if $m_i^{(p)} \neq 0$ else $0$
        \State Compute $\mathbf{M}^{(p)}$ where $\mathbf{M}^{(p)}_{i, j} \leftarrow \left(\left(\mathbf{R}_{i,j} + \sigma_i^{(p)}\right) - \sigma_i^{(p)}\right)$
        \State $\mathbf{R} = \mathbf{R} - \mathbf{M}^{(p)}$
        \EndFor
        \State $\mathbf{M}^{(s)} = \mathbf{R}$
    \end{algorithmic}
\end{algorithm}

\subsubsection{Advantages}
In addition to the advantage that the Ozaki scheme computes high-precision matrix multiplication on lower precision computing units, this scheme has three other advantages: 1) it can compute in variable precision even higher than FP64, 2) the rounding error can be smaller than the GEMMs computed by the standard floating-point arithmetics, and 3) it can be implemented by highly optimized BLAS implementation.

Regarding the first advantage, the accuracy of the Ozaki scheme can be adjusted by varying the number of splits and precision of the summing-up process for the matrix multiplication result at line 7 in Algorithm \ref{alg:ozaki-fp}.
We can improve the accuracy of matrix multiplication by increasing the number of splits and enhancing the accuracy of the summing-up process, such as using the NearSum algorithm \cite{rump_accurate_2008} to sum up the matrices in high precision.
This is also one of the advantages over the elementwise-place splitting method.
Conversely, the number of splits can be adjusted to perform matrix multiplication with an intermediate precision of IEEE standard precisions, for instance, the intermediate precision between FP32 and FP64.
This approach can improve throughput while maintaining the required accuracy.


The second advantage of the Ozaki scheme is that it can produce a smaller rounding error compared to matrix multiplication computed by standard floating-point arithmetic since the number of floating-point operations can be fewer.
In the Ozaki scheme, the computation in which rounding errors can occur is the summing-up process of the resulting matrices of each matrix multiplication (line 7 in Algorithm \ref{alg:ozaki-fp}).
The summation process is conducted $(1+s)s/2$ times for any matmul-$(m, n, k)$, a matrix multiplication of $m \times k$ matrix and $k \times n$ matrix.
For instance, the study on DGEMM on Tensor Cores by Mukunoki \textit{et al.} \cite{mukunoki_dgemm_2020} uses $s = 10 \sim 20$ splits, resulting in $55 \sim 210$ summing-up operations.
On the other hand, the number of FMA operations per one resulting element of matrix multiplication is $k$ for matmul-$(\cdot, \cdot, k)$.
This number can be larger than the number of summing-up processes in the Ozaki scheme.
Hence, the accuracy of DGEMM computation by the Ozaki scheme can be more precise than that by standard floating-point operations.

In the third advantage, we can leverage the GEMM implementations that are highly optimized for each processor, for instance, Intel MKL and NVIDIA cuBLAS, to build the Ozaki scheme implementation.
In Algorithm \ref{alg:ozaki-fp}, the matrix multiplication at line 6 is the most compute-intensive part and dominates the entire throughput.
Although we need to write the operations of the splitting (lines 1 and 2) and the accumulation (line 7) by ourselves, we can leverage the high-performance GEMM implementations and improve the entire performance.
Otherwise, similar to SGEMM emulation implementation on Tensor Cores \cite{ootomo_recovering_2022}, we would need to optimize the implementation and find the best memory-blocking parameters by ourselves.

\subsubsection{Disadvantages}
\label{sec:ozaki-disadvantage}
There are two main disadvantages of the Ozaki scheme: 1) it requires a large amount of working memory to store the matrix slices, and 2) it has weak tolerance for wide exponent range input.

The first disadvantage is a large amount of working memory to store the matrix slices.
We store all matrix slices in memory since most of the slices are used more than once.
The memory size for split matrices is linearly increased against the number of splits.
Therefore, since the Ozaki scheme improves the accuracy by increasing the number of splits, the memory limitation of the hardware is one of the limitations for the accuracy.

The second disadvantage of the Ozaki scheme is the weak tolerance for wide exponent range input.
The Ozaki scheme expands the mantissa in the mantissa space and splits them by a length $\alpha$.
When the mantissa space is wide, in other words, when the exponent distribution range of the elements is wide, we need a large number of splits to keep the entire mantissa space.
Then, a large number of splits leads to a large amount of memory consumption and computations.
And an insufficient number of splits can lead to low resulting accuracy.

\subsection{Related work}

\modified{
Mukunoki \textit{et al.} propose a method to compute the Ozaki scheme on FP16 Tensor Cores.
They split an FP64 value into FP16 values and compute the inner product on Tensor Cores by leveraging a feature of FP16 Tensor Core: the computation inside is FP32.
If we use the standard FP16 arithmetics for the computation, $k$ is limited to $2^9$ since $\alpha$ must be more than zero for $u_\text{FP16} = 2^{-11}$ in Eq. (\ref{eq:alpha}).
Although we can compute the inner product of any length of vectors by splitting every $2^9$ element and summing them up, it can not gain high performance.
On the other hand, when using Tensor Core, we can compute matrix multiplication by the Ozaki scheme where $k$ is up to $2^{22}$ since the computation inside Tensor Core is in FP32 ($u_\text{FP32}=2^{-24}$).
They demonstrated that their method outperforms the theoretical peak performance of FP64 of TITAN RTX GPU while maintaining accuracy.
}

%

\subsection{Integer matrix multiplication unit}
\begin{table}[]
\caption{The list of the architectures and processors equipped with IMMUs.}
\begin{tabular}{lll}
Arch / Processor & Input           & Output \\ \hline
NVIDIA Tensor Core \cite{nvidia_nvidia_2022}  & INT8,INT4,INT1 & INT32  \\
AMD Matrix Core \cite{amd_amd_2021}   & INT8            & INT32  \\
Intel AMX-INT8 \cite{intel_corporation_intel_2022}     & INT8            & INT32  \\
IBM POWER10 \cite{starke_ibms_2021}       & INT16,INT8,INT4 & INT32  \\
Groq TSP \cite{gwennap_groq_2020}           & INT8            & INT32  \\
ARM v8.6‑A \cite{arm_ltd_architecture_2022}           & INT8            & INT32  \\
Google TPU v1 \cite{jouppi_-datacenter_2017}      & INT8            & INT32  \\
\end{tabular}
\label{tab:imma-list}
\end{table}

The integer matrix multiplication is characterized by two properties: input and output data types.
We show the list of integer matrix multiplication architectures and processors examples in Table \ref{tab:imma-list}, where the information on the properties is available.
NVIDIA Tensor Core is mounted on a wide range of NVIDIA GPUs, from data center processors, for instance, NVIDIA A100 GPU, to edge devices, for instance, NVIDIA Jetson AGX Orin.
AMD Matrix Core is mounted on data center processors, for instance, AMD Instinct MI250 GPU.
In addition, many other integer matrix multiplication architectures and processors exist and are studied \cite{reuther_ai_2022}.
The advantages of integer arithmetics are higher throughput and power efficiency than floating-point \cite{jouppi_ten_2021,horowitz_11_2014}.

During the inference process of deep learning, we use these devices by scaling and quantizing the weights initially in floating-point values and input and output vectors.
With quantization, all values are represented as fixed-point values, and the operations can be performed as integers.
While the inference accuracy degrades by quantization, there are studies aimed at improving the accuracy \cite{jacob_quantization_2018}.

\section{DGEMM on Integer Matrix Multiplication Unit}
The block-float format stores a group of values as fixed-point values with a shared exponent.
In the Ozaki scheme, each slice of a row of matrix $\mathbf{A}$ and a column of matrix $\mathbf{B}$ can be represented in block-float format since the scheme splits the mantissa by shared exponent places.
Therefore, as the elements in a slice can be represented as fixed-point values, the computation of the Ozaki scheme is naturally suited for IMMUs.
Note that the absence of rounding errors in floating-point arithmetic in the Ozaki scheme can be seen as the absence of overflow in integer arithmetic.
This section analyzes the theoretical advantages of using IMMUs over FMMUs for the Ozaki scheme.
The summary is as follows:
\begin{enumerate}
    \item The integer method can store more valid bits per byte in a slice. Thus, we can maintain the same accuracy utilizing fewer splits than the floating-point method.
    \item The integer method uses less working memory by two factors: 1) it can reduce the duplicated exponent representation, and 2) it can reduce the number of splits.
    \item The integer method can reduce the number of matrix multiplication in the algorithm squared to the number of splits.
    \item The integer method can utilize the higher throughput of the hardware since the IMMUs typically have higher throughput than FMMUs.
    \item The INT8-input IMMU is the most suitable for the Ozaki scheme in INT32-accumulation IMMUs.
\end{enumerate}

Although the Ozaki scheme can be used for any shape of matmul-($m, n, k$), we mainly focus on $2^{11} \leq m, n, k \leq 2^{20}$ considering the effective utilization of the computing units and the memory capacity of computing devices.
We call the range as ``target range''.

\subsection{Algorithm}
\begin{algorithm}[t]
    \caption{Matrix multiplication by the Ozaki scheme on IMMU ($\odot$ is Hadamard product or elementwise product.)}\label{alg:ozaki-int}
    \begin{algorithmic}[1]
        \Require Input matrix $\mathbf{A}, \mathbf{B}$, Num split $s$
        \Ensure $\mathbf{C} \leftarrow \mathbf{A}\cdot\mathbf{B}$
        \State $\left(\mathbf{A}^{(1)}, \mathbf{A}^{(2)}, \cdots, \mathbf{A}^{(s)}\right), \mathbf{e}_A\leftarrow \text{SplitInt}\left(\mathbf{A}, s\right)$
        \State $\left(\mathbf{B}^{(1)}, \mathbf{B}^{(2)}, \cdots, \mathbf{B}^{(s)}\right), \mathbf{e}_B\leftarrow \text{SplitInt}\left(\mathbf{B}, s\right)$
        \State $\mathbf{C} = 0$
        \For $\text{ \ }i = 1..s$
        \For $\text{ \ }j = 1..(s-i+1)$
        \State $\mathbf{C}_\text{tmp} \leftarrow \mathbf{A}^{(i)}\cdot\mathbf{B}^{(j)}$ // Integer
        \State $\mathbf{C} \leftarrow \mathbf{C} + \mathbf{C}_\text{tmp} \odot \left(2^{-(i + j)\alpha}\cdot \mathbf{e}_A \cdot {\mathbf{e}_B}^\top\right)$
        \EndFor
        \EndFor
    \end{algorithmic}
\end{algorithm}

\begin{algorithm}[t]
    \caption{SplitInt: Splitting algorithm of the Ozaki scheme by integer arithmetics for $m \times k$ input matrix $\mathbf{M}=\mathbf{A} \text{ or } \mathbf{B}^\top$}\label{alg:ozaki-int-splitting}
    \begin{algorithmic}[1]
        \Require Input $m \times k$ matrix $\mathbf{M}$, Num split $s$
        \Ensure Matrix slices : $\mathbf{M}^{(1)}, \mathbf{M}^{(2)}, \cdots, \mathbf{M}^{(s)}$, $m$-dimensional exponent offset vector $\mathbf{e}$
        \State $\alpha = \lfloor(-\log_2 u + \log_2 k) / 2\rfloor$
        \State Compute $\mathbf{e}$ where $\mathbf{e}_i \leftarrow \max_j2^{\lceil\log_2|\mathbf{R}_{i,j}|\rceil}$
        \For $\text{ \ }p = 1..s$
        \State Compute $\mathbf{M}^{(p)}$ where
        \begin{itemize}
            \item the sign of $\mathbf{M}^{(p)}_{i, j}$ is the same as $\mathbf{M}_{i, j}$
            \item the mantissa of $\mathbf{M}^{(p)}_{i, j}$ is part of $\mathbf{M}_{i, j}$ mantissa, which corresponds to the part from $2^{(p-1)\alpha}$ bit to $2^{p\alpha}-1$ bit in the mantissa space with $\mathbf{e}_i$ as the first bit, as shown at the right of Fig. \ref{fig:ozaki-scheme}.
        \end{itemize}
        \EndFor
    \end{algorithmic}
\end{algorithm}

\begin{figure}
    \centering
    \includegraphics[width=\linewidth]{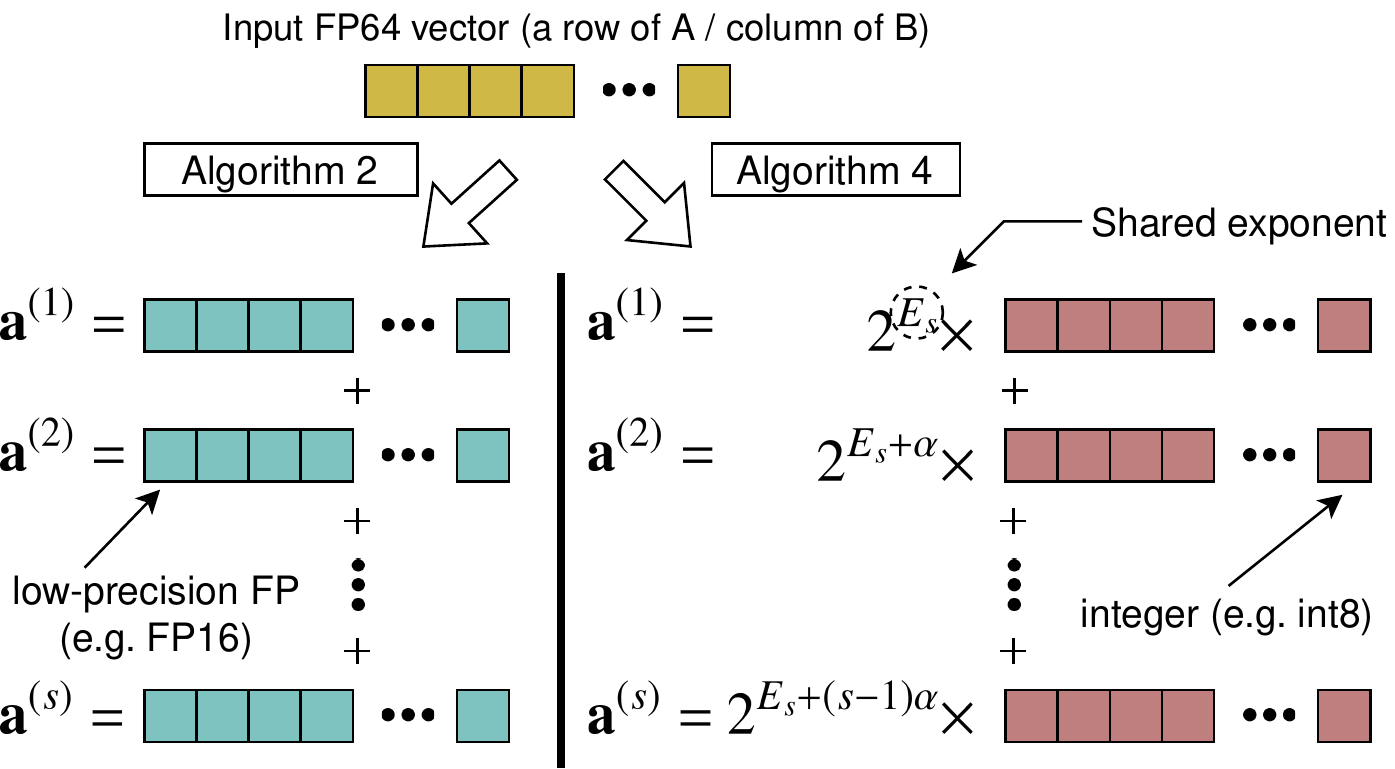}
    \caption{The difference of the data storing ways in Algorithm \ref{alg:ozaki-fp-splitting} (Original algorithm using floating-point) and Algorithm \ref{alg:ozaki-int-splitting} (Using integer).}
    \label{fig:algo-diff}
\end{figure}

We show the Ozaki scheme and its splitting algorithm for integer arithmetics in Algorithm \ref{alg:ozaki-int} and \ref{alg:ozaki-int-splitting}.
Although the basic operations are the same as Algorithm \ref{alg:ozaki-fp} and \ref{alg:ozaki-fp-splitting}, we use integer arithmetic operations instead of floating-point arithmetic operations.
The error analysis of the Ozaki scheme on IMMU is the same as the floating-point operation \cite{ozaki_error-free_2012} since the method itself is the same.
We show the difference in storing format of the slices between Algorithm \ref{alg:ozaki-fp-splitting} and \ref{alg:ozaki-int-splitting} in Fig. \ref{fig:algo-diff}.

\subsection{Advantages}
\label{sec:imma-advantage}

\begin{table}[t]
    \caption{
    The list of matrix multiplication unit specifications to consider and compare for use in the Ozaki scheme.
    The label is \{Input data type\}-\{Output/accumulation data type\}.
    }
\begin{tabular}{l|cc|c}
            & \multicolumn{2}{c|}{mantissa length {[}bit{]}}              & \multirow{2}{*}{\begin{tabular}[c]{@{}c@{}}input data\\ size {[}Byte{]}\end{tabular}} \\
IMMU/FMMU   & input ($\ell_\text{in}$) & accumulation ($\ell_\text{acc}$) &                                                                                       \\ \hline
FP16-FP32   & 11                       & 24                               & 2                                                                                     \\
INT4-INT32  & 3                        & 31                               & 0.5                                                                                   \\
INT8-INT32  & 7                        & 31                               & 1                                                                                     \\
INT12-INT32 & 11                       & 31                               & 1.5
    \end{tabular}
    \label{tab:mau}
\end{table}

\begin{figure}[t]
    \centering
    \includegraphics[width=\linewidth]{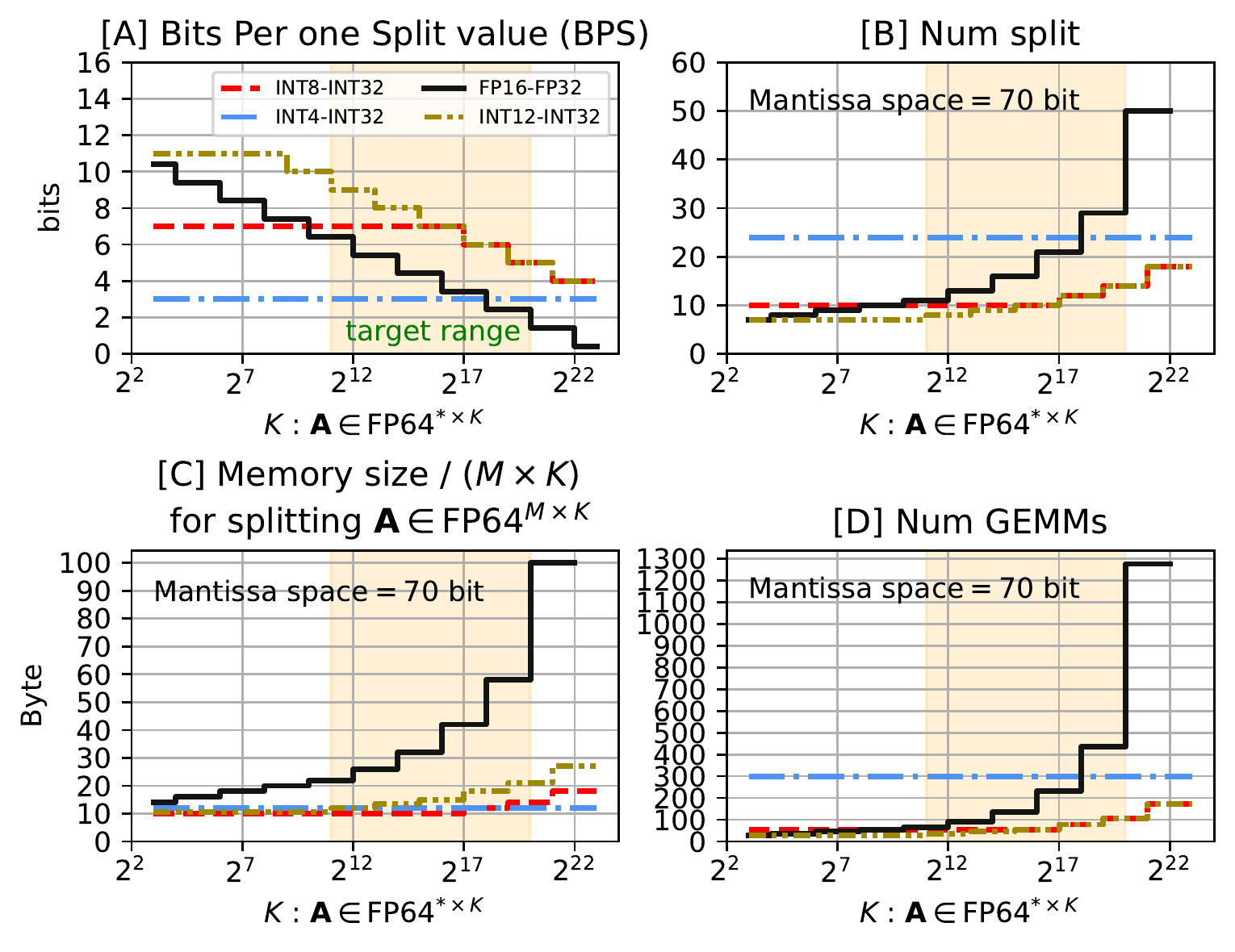}
    \caption{
    Comparing memory consumption and the number of GEMM operations among the matrix multiplication units.
    \textit{Upper left:} The mantissa bit length one slice value keeps (BPS).
    \textit{Upper right:} The number of splits to keep a specific mantissa space length (representation accuracy).
    The mantissa space length is $\text{BPS} \times \text{num\_split}$.
    \textit{Bottom left:} The working memory size for storing the slices to keep a specific mantissa space length.
    \textit{Bottom right:} The number of the matrix multiplications at line 6 in Algorithm \ref{alg:ozaki-fp} and \ref{alg:ozaki-int}.
    }
    \label{fig:int-ozaki-shceme}
\end{figure}

\subsubsection{Larger bits per slice (BPS)}
\label{sec:pbs}
The mantissa bit length of a slice, $\alpha$, depends on the precision of the accumulation and the number of accumulations, as described by Eq. (\ref{eq:alpha}).
We show the number of mantissa bits per split (BPS) at the top left of Fig. \ref{fig:int-ozaki-shceme} for each matrix multiplication unit in Table \ref{tab:mau}.
We derive the Eq. (\ref{eq:alpha}) for the mantissa bit length and obtain the BPS as follows:
\begin{align}
\alpha &= \lfloor (\ell_\text{acc} - \log_2 k) /2 \rfloor\\
\text{BPS} &= \text{min}\left(\alpha, \ell_\text{in}\right).
\end{align}
When using FP16-FP32, the BPS decreases by increasing the accumulation size $k$ since the mantissa length of the accumulator is relatively small, and the BPS equals $\alpha$.
In this case, $\alpha$ is smaller than $\ell_\text{in}$.
In other words, each FP16 mantissa in the slices is not filled effectively, and waste bits exist, which is calculated as $\ell_m - \text{BPS}$.
Using INT12-INT32 results in a larger BPS compared to FP16-FP32 since the accumulator mantissa length ($\ell_\text{acc}$) is larger while $\ell_\text{in}$ is the same.
However, there are still wasted bits of up to $5$ bits in the target range.
Using INT8-INT32 results in a BPS that is equal to $\ell_\text{in}$ for $k < 2^{18}$ and equal to $\alpha$ for $k \geq 2^{18}$, which is larger than the FP16-FP32 BPS.
Although there are still wasted bits, they are up to $1$ bit in the target range.
Using INT4-INT32 results in a BPS that is equal to $\ell_\text{in}$, and no wasted bits exist.
However, the BPS is smaller than FP16-FP32 for most of the target range.

\subsubsection{Fewer number of splits to preserve the same mantissa space}
\label{sec:fewer-num-splits}
The number of splits determines the trade-off between accuracy, the number of arithmetic operations, and memory consumption.
Specifically, the accuracy is the length of the mantissa space the entire slices keep, calculated as ``the number of splits'' $\times$ BPS.
We show the number of splits required to maintain a certain mantissa space length, for instance, $70$, at the top right of Fig. \ref{fig:int-ozaki-shceme}.
Each line is inversely proportional to the BPS.
In most cases in the target range, the number of splits of INT8-INT32 and INT12-INT32 is smaller than FP16-FP32.
On the other hand, INT4-INT32 is larger than FP16-FP32.
Although we show only one example of mantissa space length in the figure, this trend is the same with other mantissa space lengths.
Therefore, we believe that discussing the advantages using this example mantissa space length represents the general trend.

\subsubsection{Smaller memory size for keeping the slices}
We explicitly keep the slices of the input matrices on memory to reuse the data.
The required memory size is proportional to ``the number of splits'' $\times$ ``storage size of the slice.''
The storage data type is the same as the input data type of a matrix multiplication unit, and its size is shown in Table \ref{tab:mau}.
We show the memory size per one element at the bottom left of Fig. \ref{fig:int-ozaki-shceme}.
The memory size when using IMMU is smaller than FP16-FP32 FMMU.
This is due to the following two reasons:
\begin{enumerate}
    \item We can reduce the number of splits when using INT8-INT32 and INT12-INT32, as mention in Sec. \ref{sec:fewer-num-splits}.
    \item By using integers to store the slices, we can reduce the duplicated exponent information that occurs when using FP16-FP32.
    When storing the slices as floating-point values, each element of each slice has its own exponent.
    In contrast, with integer storage, we only need to store one shared exponent in every element of every slice, as illustrated in Figure \ref{fig:algo-diff}.
    This way, we can effectively reduce the amount of memory required to store each exponent and focus on storing the mantissa part.
\end{enumerate}
Among the IMMUs, INT8-INT32 consumes the least memory since it generates fewer waste bits than INT12-INT32, as discussed in Section \ref{sec:pbs}.
Additionally, INT8-INT32 offers a higher mantissa information density (mantissa bit length per data size) compared to INT4-INT32.

\subsubsection{Fewer number of GEMM operations}
The number of GEMM operations (line 6 in Algorithm \ref{alg:ozaki-fp} and \ref{alg:ozaki-int}) is $(s + 1)\times s / 2$, where $s$ is the number of splits.
We show the number of operations at the bottom right of Fig. \ref{fig:int-ozaki-shceme}.
INT12-INT32 requires the fewest number of operations in the target range.
Although INT8-INT32 requires about 1.52 times more operations than INT12-INT32 for $k < 2^{16}$, the throughput advantage of INT12-INT32 over INT8-INT32 depends on the hardware implementation since higher precision computing units typically have a slower throughput.
For instance, in the case of NVIDIA Tensor Cores, the throughput of INT8 Tensor Core (INT8-INT32) is half of INT4 Tensor Core (INT4-INT32).
Therefore, the throughput advantage of INT12-INT32 against INT8-INT32 depends on the hardware implementation.
In $m \geq 2^{16}$, both INT8-INT32 and INT12-INT32 require the same number of operations, and their number of operations is smaller than that of FP16-FP32.
In contrast, INT4-INT32 requires about six times more operations than the other IMMUs.
In the case of NVIDIA GPUs, the throughput of the Ozaki scheme on INT4 Tensor Core can be slower than that on INT8 Tensor Cores since the throughput of INT4 Tensor Core is only two times faster than that of INT8.

\subsection{Disadvantages}
There is no theoretical disadvantage in using fixed-point arithmetic and IMMU in the Ozaki scheme since the scheme implicitly uses fixed-point values to avoid rounding errors even in the computation on floating point computing units.

\subsection{Suitable IMMU for the Ozaki scheme}
The memory usage and the number of GEMM operations are affected by the properties of the IMMU, specifically the mantissa length of inputs and accumulation.
If the mantissa bit length of the input integer is too large, it can result in waste bits and increase the required memory size.
On the other hand, if the mantissa bit length is too short, the number of splits required to keep the mantissa space increases, which in turn increases the number of computations. 
While the suitable IMMU depends on the size of matrix multiplication, we believe that INT8-INT32 is the most suitable for the target range, considering the theoretical advantages discussed in Section \ref{sec:imma-advantage}.

\section{Experiments}
We utilized NVIDIA GPUs for our experiments due to their easy availability and the maturity of the software development environment within the processors that have IMMUs.

\subsection{Preparation}
\begin{figure}[t]
    \centering
    \includegraphics[width=\linewidth]{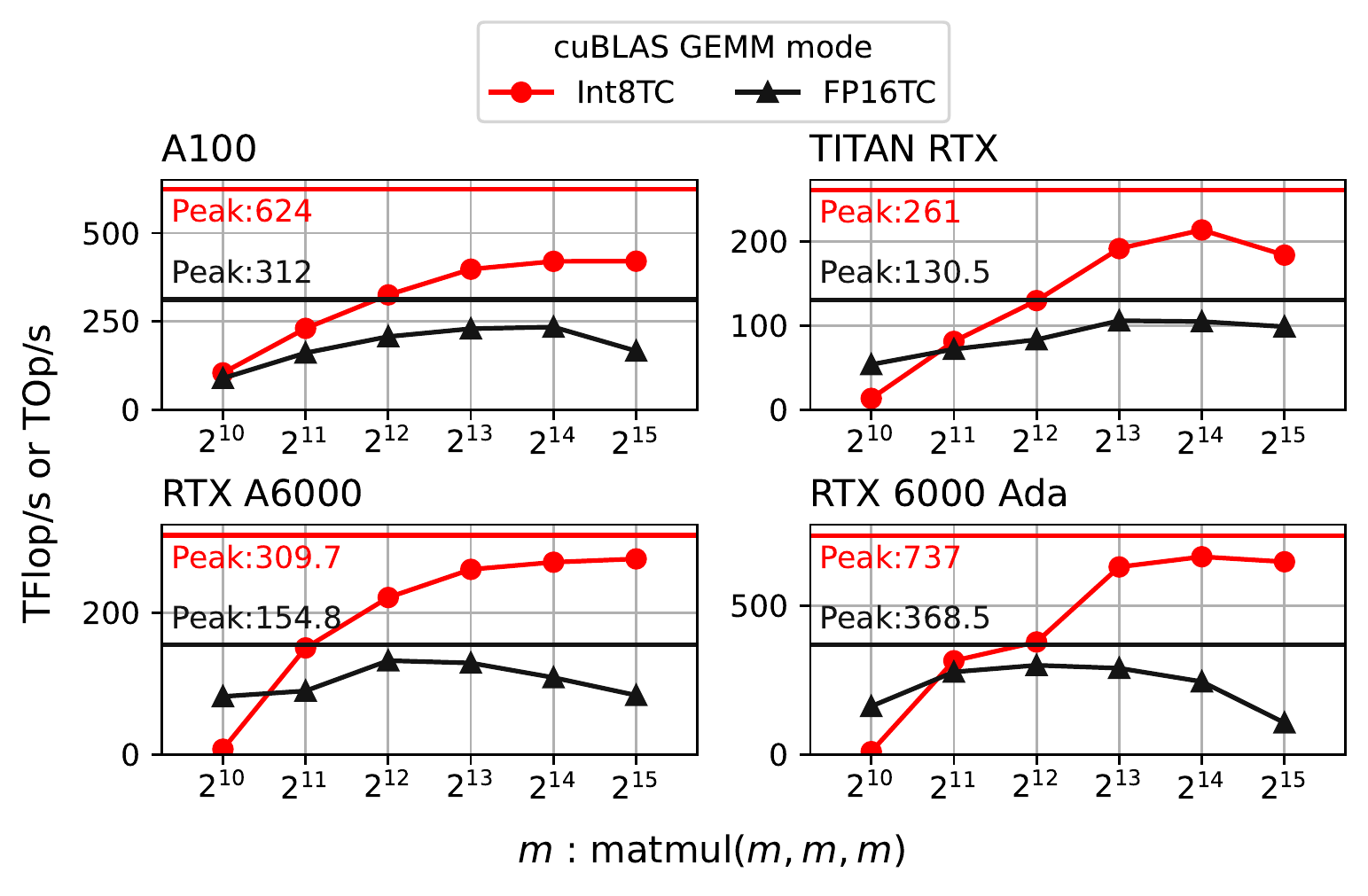}
    \caption{The unit throughput comparison between INT8 (INT8-INT32) and FP16 Tensor Core (FP16-FP32).}
    \label{fig:gemm-unit-test}
\end{figure}
We have implemented a matrix multiplication library, ozIMMU \footnote{\url{https://github.com/enp1s0/ozimmu}}, for NVIDIA GPUs using the NVIDIA cuBLAS library and custom kernel functions.
Our library uses bit operations to cut the mantissa and make a series of matrix slices.
Subsequently, we compute each matrix multiplication $\mathbf{A}^{(i)} \cdot \mathbf{B}^{(j)}$ in Algorithm \ref{alg:ozaki-int} using the cuBLAS function {\tt cublasGemmEx}.
The label of our implementation is "{\tt INT8x}$X$", where $X$ is the number of splits $s$.

To ensure that the throughput of INT8 (INT8-INT32) Tensor Cores is faster than FP16 Tensor Core (FP16-FP32) in practice, we conducted a preliminary experiment to compare them, as depicted in Fig. \ref{fig:gemm-unit-test}.
Although Tensor Core supports INT4 input matrices, the cuBLAS function does not provide the interface, and therefore we do not evaluate its throughput.
We have confirmed that the throughput of INT8 Tensor Cores is higher than FP16 Tensor Cores when the size of the input matrice is large enough.

\subsection{Accuracy}
We conduct two numerical experiments to evaluate accuracy: 1) exponent distribution range tolerance and 2) zero-cancellation evaluation.
\subsubsection{exponent distribution range tolerance evaluation}
\begin{figure}[t]
    \centering
    \includegraphics[width=\linewidth]{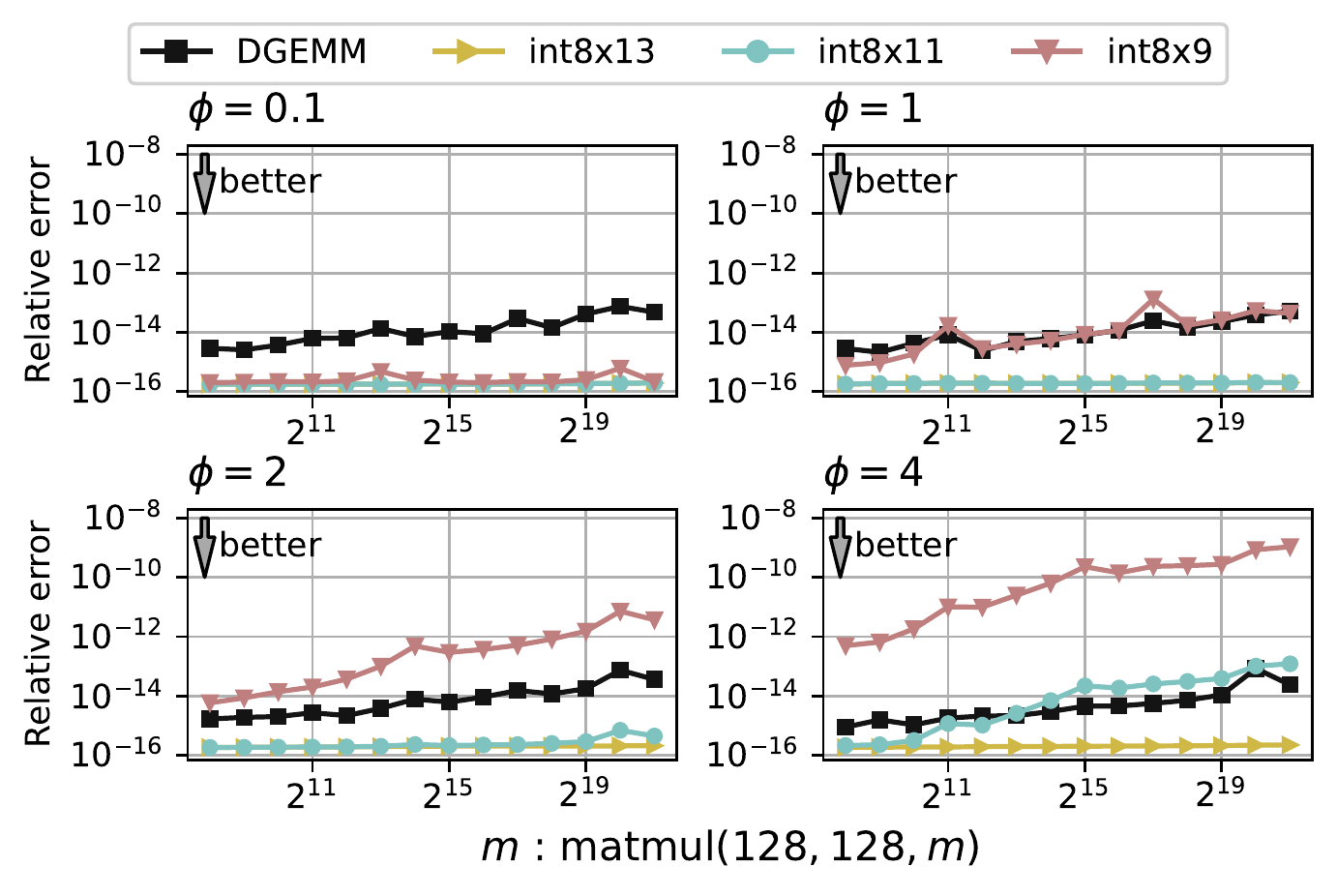}
    \caption{The accuracy evaluation of matrix multiplications where the elements of input matrices have various exponent ranges ($\phi=0.1, 1, 2$ and $4$).}
    \label{fig:exponent-accuracy}
\end{figure}
As mentioned in Sec. \ref{sec:ozaki-disadvantage}, one of the disadvantages of the Ozaki scheme is the need to increase the number of splits to maintain accuracy when the exponent distribution of the matrix elements is large.
To investigate the effect of the exponent distribution on accuracy, we measured the error of matrix multiplications $\mathbf{A}\cdot\mathbf{B}$ for different exponent distribution random inputs.
Each element of the input matrix is generated as follows:
\begin{equation}
    \mathbf{A}_{i, j}, \mathbf{B}_{i, j} = \text{uniform}(-0.5, 0.5) \times e^{\phi \times \text{normal(0, 1)}},
\end{equation}
where $\phi$ is a parameter to control the exponent distribution range.
This generation method is used in \cite{ozaki_error-free_2012,mukunoki_dgemm_2020}.
We measured the error using the average relative error of resulting matrix elements, where the relative error is calculated as follows:
\begin{equation}
    \label{eq:relative-error}
    \text{relative \ error}_{i,j} = |\mathbf{C}_{i, j} - \mathbf{C}_{i, j}^\text{DD}|/|\mathbf{C}_{i, j}^\text{DD}|,
\end{equation}
$\mathbf{C}$ is the resulting matrix, and $\mathbf{C}^\text{DD}$ is the reference resulting matrix computed in higher precision, the {\tt double-double} precision.
We show the results for $\phi=0.1, 1, 2$ and $4$ in Fig. \ref{fig:exponent-accuracy}.
While the error of {\tt INT8x}$9$ is smaller than DGEMM when the exponent distribution is narrow ($\phi = 0.1$), the error becomes large as the exponent range extends from $\phi=1$ to $4$.
In contrast, for {\tt INT8x}$11$ and {\tt INT8x}$13$, the error is either smaller or almost at the same level as DGEMM, even when the exponent distribution range is large ($\phi=4$).
We also calculated the maximum value of the related errors and confirmed that the trend is consistent with the average.

\subsubsection{zero-cancellation evaluation}
\begin{figure}[t]
    \centering
    \includegraphics[width=\linewidth]{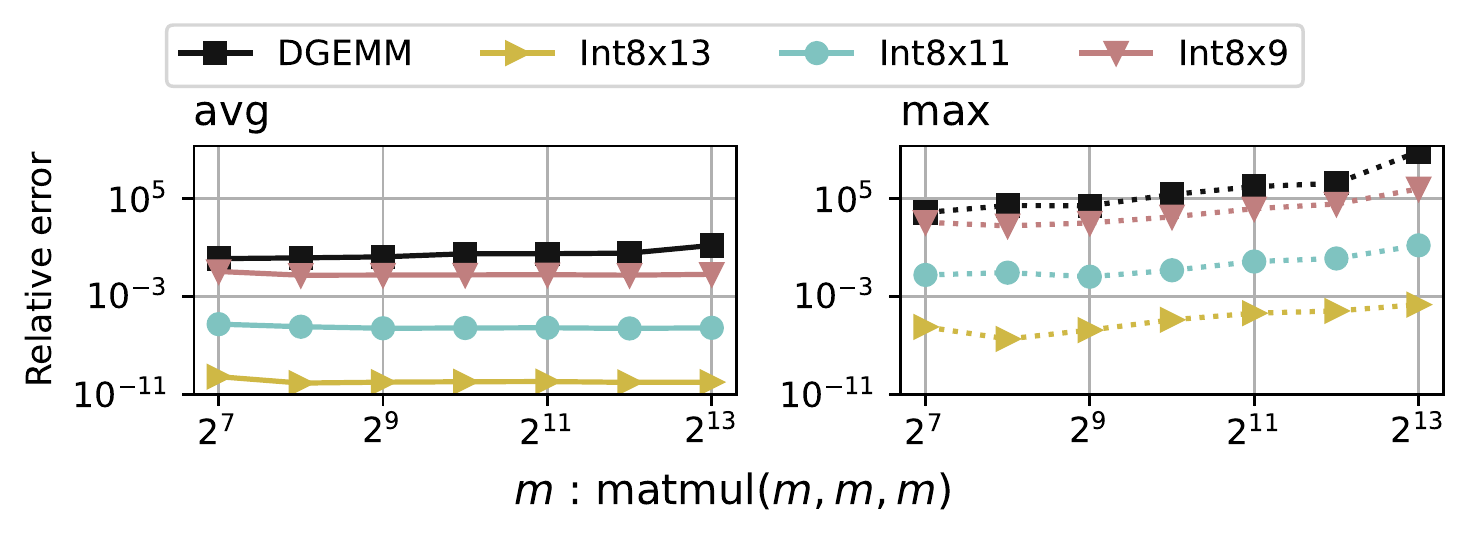}
    \caption{The accuracy comparison of matrix multiplication of a matrix and its approximate inverse $\mathbf{A}\cdot\mathbf{A}^{\dag}$.}
    \label{fig:accuracy-invpair}
\end{figure}
\modified{Multiplying a matrix $\mathbf{A}$ and its approximate inverse (or right inverse) $\mathbf{A}^{\dag}$ leads to zero off-diagonal elements in the resulting matrix.
This computation is sensitive to rounding errors, which can affect the accuracy significantly.
To evaluate the error of the multiplication $\mathbf{A}\cdot\mathbf{A}^{\dag}$, we generate a random matrix $\mathbf{A}$ with a normal distribution of $\mathcal{N}(0, 1)$ and compute $\mathbf{A}^{\dag}$ by solving the linear equation $\mathbf{A}\cdot\mathbf{X}=\mathbf{I}$, where $\mathbf{I}$ is an identity matrix.
The results are shown in Fig. \ref{fig:accuracy-invpair}.
Note that we use the resulting reference matrix in Eq. (\ref{eq:relative-error}) calculated in {\tt double-double} precision, not the ideal identity matrix.
The error of {\tt INT8x}$X$ is smaller than DGEMM since the Ozaki scheme calculates the cancellation of the high digit part of the resulting mantissa with higher accuracy.
When using floating-point arithmetic, the subtraction of two close values leads to the cancellation of mantissa digits, resulting in accuracy loss even in the high digits of the mantissa.
In contrast, when using the Ozaki scheme, the resulting mantissa is computed digit block by digit block from the MSB of the mantissa without accuracy loss, allowing us to compute high digits of the mantissa with high accuracy.
We can improve the accuracy further by splitting the input into more digit blocks.}

\subsection{Throughput and power efficiency}
\begin{figure*}[t]
    \centering
    \includegraphics[width=\textwidth]{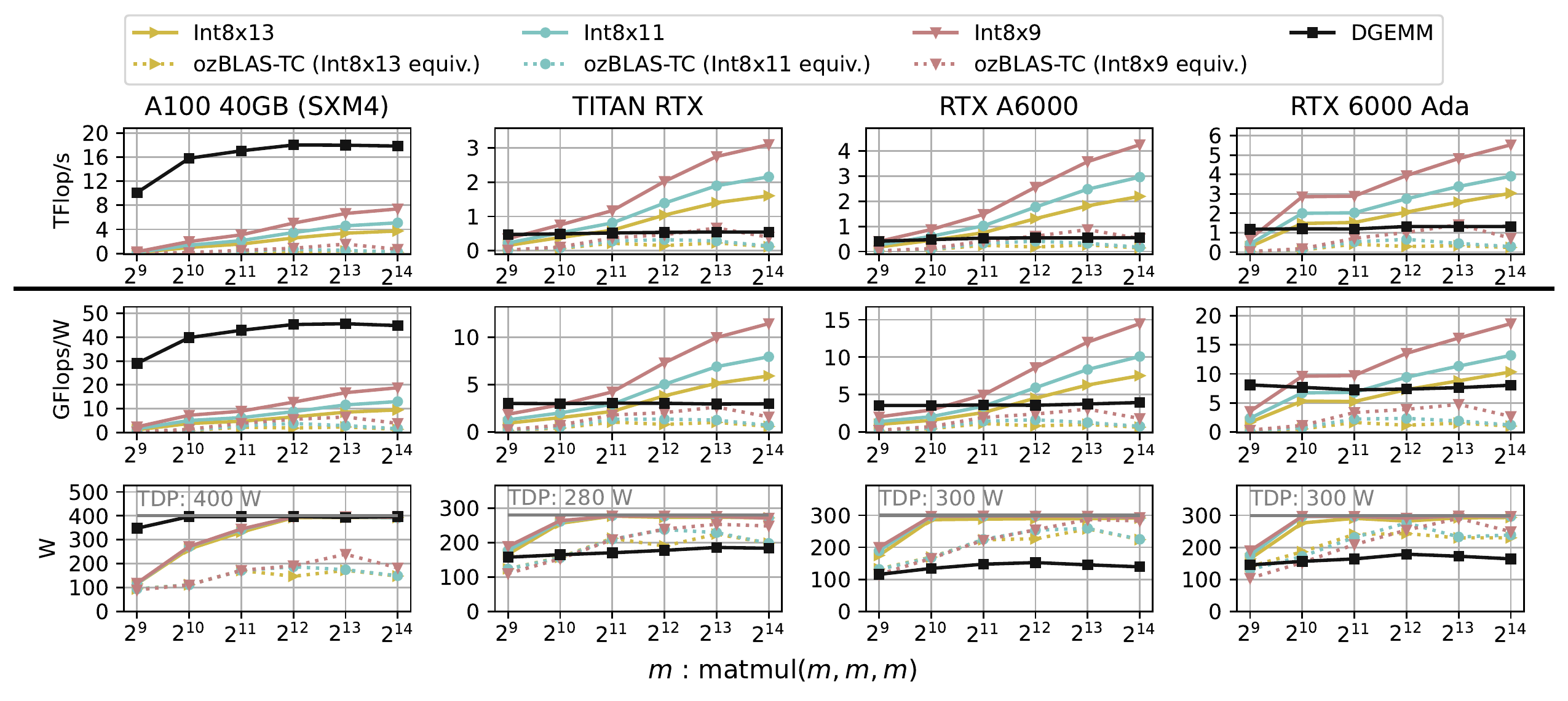}
    \caption{
    The throughput (top row), power efficiency (middle row), and power consumption (bottom row) comparison.
    The {\tt INT8x}$X$ implementation is the Ozaki scheme on integer Tensor Cores, and ozBLAS is the implementation on FP16 Tensor Cores by Mukunoki \textit{et al.}
    We select the number of splits of the ozBLAS implementation so that the mantissa space length is the same as {\tt INT8x}$X$, and the accuracy kept by the slices is almost equivalent.
    }
    \label{fig:throughput}
\end{figure*}
In our evaluation of throughput, we measure the computation time $t$ [s] for matmul-$(m, n, k)$ and compute the throughput $2mnk/t$ [Flop/s].
The {\tt INT8x}$X$ implementations calculate the matrix multiplication by integer operations, and in addition, the accuracy is not identical for DGEMM depending on the exponent distribution of the input matrices.
Nonetheless, we consider that this indicator is helpful to compare the throughput among the implementations and evaluate the trade-off between the accuracy and throughput.
We evaluate the throughput on four GPUs: NVIDIA A100, TITAN RTX (Turing architecture), RTX A6000 (Ampere architecture), and RTX 6000 Ada (Ada Lovelace architecture).
The results are shown in the top row of Fig. \ref{fig:throughput}.
On A100, DGEMM achieves a throughput of over 90\% of the theoretical peak performance of FP64 Tensor Cores ($19.5$ TFlop/s), while {\tt INT8x}$X$s are $3 \sim 5 \times$ slower than DGEMM.
This is expected since the number of integer matrix multiplication operations inside {\tt INT8x}$X$s is $45 \sim 108$, whereas the max performance of integer Tensor Cores is only up to $20$ times faster than FP64 Tensor Cores.
In contrast, on the other GPUs with low FP64 theoretical peak performance, {\tt INT8x}$X$s are faster than DGEMM.
We also compare the Ozaki scheme on integer Tensor Cores to the one on FP16 Tensor Cores by Mukunoki \textit{et al.} \cite{mukunoki_dgemm_2020}, which is available online\footnote{\url{https://www.r-ccs.riken.jp/labs/lpnctrt/projects/gemmtc/index.html}. We use version 1.3a.}.
In all cases and on all GPUs, the integer Tensor Core implementations are faster than the FP16 ones.

In the power efficiency evaluation, we obtain the total power consumption [W$\cdot$s] during the GEMM computation and calculate the efficiency [Flop/W].
To obtain the power consumption, we run the GEMM operation for at least 10 seconds and measure the power consumption [W] at 100ms intervals using NVML (NVIDIA Management Library) .
We show the power efficiency [GFlops/W] and the average power consumption [W] in the middle and bottom of Fig. \ref{fig:throughput}.
Our results show that, across all GPUs, the average power consumption of {\tt INT8x}$X$ is almost equivalent to the device's TDP (Thermal Design Power).
Therefore, the power efficiency is proportional to the throughput when the matrix size is large.
On the other hand, although the average power consumption of DGEMM is almost equivalent to the device's TDP on the NVIDIA A100, it is smaller on the other GPUs.
Although the power efficiency of {\tt INT8x}$X$ is lower than DGEMM on A100, it is higher on the other consumer GPUs.

\begin{figure}
    \centering
    \includegraphics[width=\linewidth]{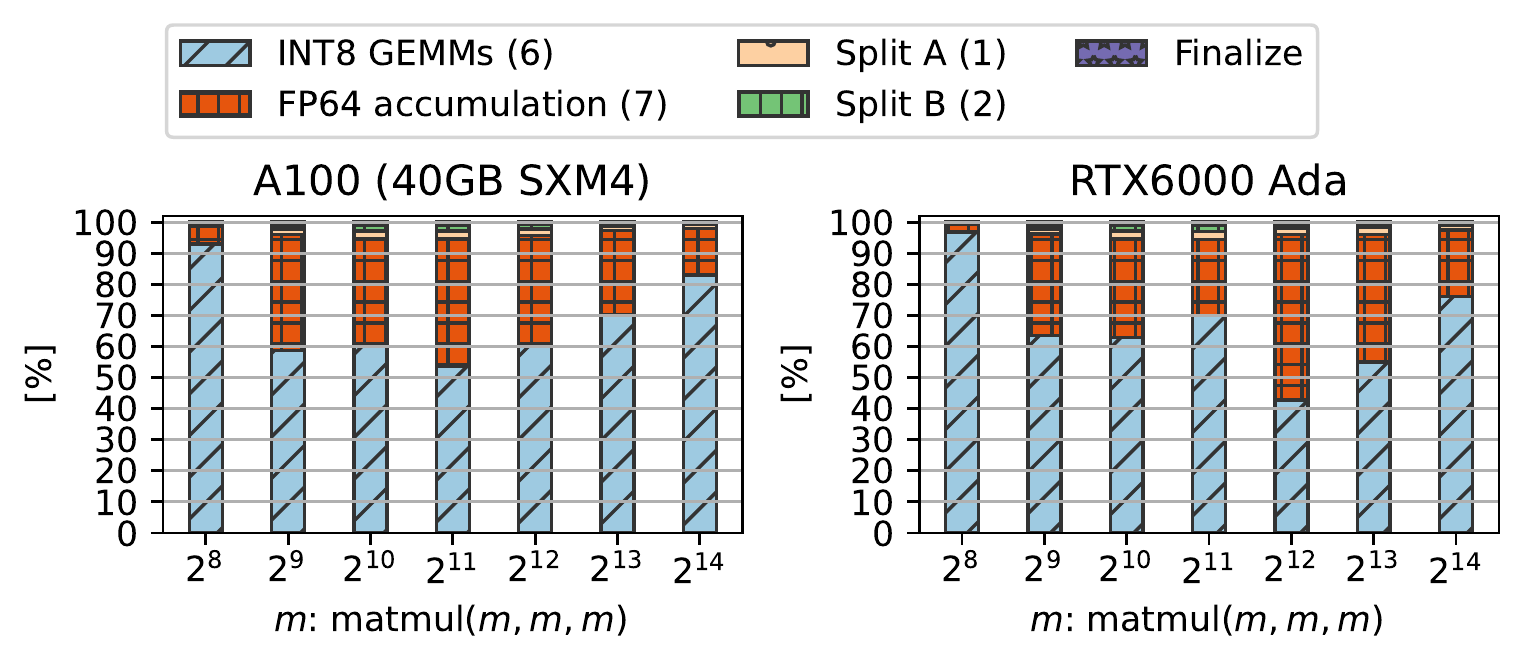}
    \caption{The time breakdown of INT8x9. The number in the parenthesis in the legend is the line number of Algorithm \ref{alg:ozaki-int}.}
    \label{fig:int8x9-breakdown}
\end{figure}
For a more detailed throughput analysis, we present the time breakdown of {\tt INT8x}$9$ on NVIDIA A100 and RTX 6000 Ada GPUs in Fig. \ref{fig:int8x9-breakdown}.
We observe that the INT8 GEMM computations and the FP64 accumulation of the resulting matrix primarily consume the computation time.
We further analyzed the performance bottleneck of the FP64 accumulation function by NVIDIA Nsight Compute.
We found that for problem sizes $m \leq 2^{11}$, the kernel function does not fully utilize the computing resources of the GPUs, leading to suboptimal performance.
However, for problem sizes $m \geq 2^{12}$, the kernel function utilizes approximately 90\% of the bandwidth of the device memory on both GPUs.
Therefore, to improve the throughput further, modifying the algorithm and reducing the number of FP64 accumulations is necessary.
The ratio of FP64 accumulation in our implementation is higher than that of the implementation on FP16 Tensor Cores by Mukunoki \textit{et al.} \cite{mukunoki_dgemm_2020} because integer Tensor Core achieves faster throughput than FP16.

\subsubsection{Comparison to other high precision GEMM method}
To the extent of our knowledge, there is currently no efficient implementation for double-precision equivalent matrix multiplication using lower-precision arithmetics on GPUs.
One method that can be used to compute matrix multiplication with higher precision than single-precision is the single-single approach, which employs two FP32s for one element.
Despite the total mantissa length of two FP32s being smaller than that of FP64, we calculate the theoretical peak throughput of single-single GEMM and compare it to the throughput of the Ozaki scheme on IMMU.
To perform a multiply-and-add (MAD) operation for two single-single values, a certain number of single-precision instructions are executed.
We need $11$ add instructions for addition, $4$ multiply instructions, $9$ add instructions, and $5$ FMA instructions for multiplication \cite{nakata_mplapack_2022}\footnote{We have counted the number of instructions in this implementation.}.
We assume that all additions and multiplications are executed as FMA instructions since the theoretical peak throughput of a GPU is calculated based on FMA instruction throughput.
By calculating the number of FMA instructions required to compute MAD, which is $39$, we determine that the theoretical peak throughput of single-single MAD is $1/(29 \times 2)$ of the theoretical peak throughput of the single-precision computing unit.
This implies that the throughput of single-single precision matrix multiplication is almost equivalent to or slower than that of double-precision computing units.
Therefore, we conclude that there is no throughput advantage in using single-single precision matrix multiplication compared to the Ozaki scheme on IMMU or double-precision Tensor Cores on A100 GPU.

\subsection{Application: Quantum Circuit Simulation}
\begin{figure}
    \centering
    \includegraphics[width=\linewidth]{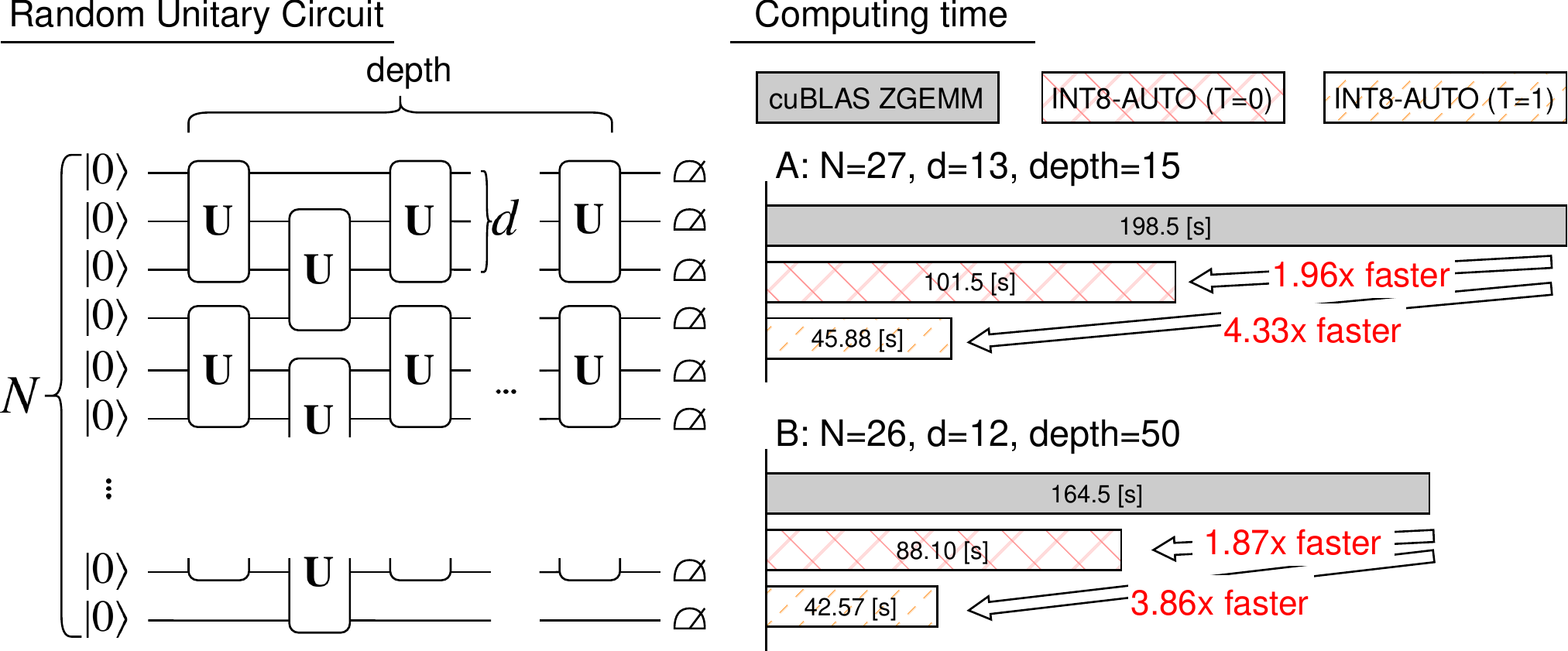}
    \caption{\textit{Left:} The structure of brickwork random unitary circuit. \textit{Right:} The computing time comparison of a StateVec simulation using cuBLAS DGEMM and the Ozaki scheme on integer Tensor Cores.}
    \label{fig:unitary-quantum}
\end{figure}
We apply the Ozaki scheme on integer Tensor Cores to a state vector (StateVec) quantum circuit simulation.
In the StateVec simulation, first, we prepare a $2^N$ size of a double-precision complex array (state vector) on memory, where $N$ is the number of qubits.
Then we update the values in the state vector by the operation determined by a quantum gate.
The computation of applying a quantum gate is 1) reshaping the state vector to a matrix and 2) multiplying a unitary matrix representing the quantum gate.
In the case of $d$-qubit gate, the shape of the matrix multiplication is matmul-$(2^{N-d}, 2^d, 2^d)$.
When $d$ is large enough, this matrix multiplication is compute-intensive, and there is a motivation to improve its throughput.
To apply the Ozaki scheme on integer Tensor Cores to the simulation, we extend the implementation as follows:
\begin{itemize}
    \item Supporting complex GEMM.
    We can compute a complex GEMM by separating the real and imaginary parts and computing a series of GEMMs among them.
    We separate them while splitting the mantissa in the first phase of the Ozaki scheme.
    \item Supporting automatic selection of the number of splits ({\tt INT8}-AUTO).
    We can not determine the number of splits before starting the simulation since we do not know the exponent distribution range of the input matrices during the simulation.
    Therefore, before a GEMM computation, we check all the elements of the input matrices and determine the appropriate number of splits.
    We select the number of splits so that the average mantissa loss in the splitting process is equal to or smaller than a threshold $T$.
    For instance, when we set $T = 0$, no mantissa loss occurs in the splitting operation.
    \item Intercepting cuBLAS double-precision GEMM function calls and executing {\tt INT8}-AUTO instead.
    We use an environmental variable {\tt LD\_PRELOAD} to realize it.
\end{itemize}

In this evaluation, we simulate a brickwork random unitary circuit, which is used to investigate the universal properties of quantum systems and understand quantum dynamics.
The random unitary circuit is used as a toy model of the systems, including quantum chaos \cite{nahum_operator_2018}, gravity theory \cite{hayden_black_2007}, and quantum error correction code \cite{choi_quantum_2020} that lack specific structures, including symmetries, which make exact analysis impractical \cite{fisher_random_2023}.
We apply several $d$-qubit unitary gates $\mathbf{U}$ to the initial states of the qubits, as shown at the left of Fig. \ref{fig:unitary-quantum}.
Each unitary matrix is generated by a QR decomposition of a Gaussian random complex matrix.
We compare the throughput and accuracy of the simulation among by cuBLAS ZGEMM, {\tt INT8-AUTO} ($T=0$), and {\tt INT8-AUTO} ($T=1$) for two circuit configurations, A and B.
The configurations and the average throughput of ten executions on NVIDIA RTX6000 Ada GPU are shown at the right of Fig. \ref{fig:unitary-quantum}.
When not admitting the loss of mantissa during the splitting process ($T=0$), the {\tt INT8x12} and {\tt INT8x13} modes are automatically selected, and the speed-up ratio is $1.96$ in A and $1.87$ in B.
On the other hand, when we admit a 1-bit loss of mantissa on average in the splitting process ($T=1$), the {\tt INT8x8} and {\tt INT8x9} modes are automatically selected, and the speed-up ratio is $4.33$ in A and $3.86$ in B.
For the accuracy evaluation, we compare the relative error of the real part of the first element of the state vector (the amplitude of $|00...0\rangle$) to a reference value computed in double-double precision.
\begin{table}[]
\setlength{\tabcolsep}{3pt}
\caption{
The relative error and memory consumption in the simulation of random unitary circuit simulation.
The memory consumption is the amount of memory to store the splits in the Ozaki scheme.
}
\label{tab:random-unitary-result}
\begin{tabular}{l|lll|ll}
  & \multicolumn{3}{l|}{Average relative error}                         & \multicolumn{2}{l}{Memory {[}GB{]}} \\
  & cuBLAS                & AUTO ($T=0$)            & AUTO ($T=1$)           & $T=0$           & $T=1$           \\ \hline
A & $1.3\times 10^{-13}$ & $1.4 \times 10^{-13}$ & $1.3\times 10^{-13}$ & 7.9                  & 6.3                  \\
B & $2.8\times 10^{-13}$ & $3.0 \times 10^{-13}$ & $3.9\times 10^{-13}$ & 3.5                  & 2.7                 
\end{tabular}
\end{table}
The average relative errors of cuBLAS ZGEMM, INT8-AUTO ($T=0$), and INT8-AUTO ($T=1$) are shown in Table \ref{tab:random-unitary-result}.
We consider that the error of the Ozaki scheme on integer Tensor Cores is almost at the same level as cuBLAS ZGEMM.
Furthermore, we also measure the memory consumption during the simulation to store the splices in the Ozaki scheme, as shown in Table \ref{tab:random-unitary-result}.
Although the Ozaki scheme on integer Tensor Cores has consumed several GB of memory, this is about 50\% less than FMMA.
In summary, the Ozaki scheme on integer Tensor Cores has computed the simulation up to $4.33$ times faster than cuBLAS ZGEMM while the accuracy is almost at the same level.

\section{Conclusion}
In conclusion, this paper presents the theoretical advantages and practical benefits of using the Ozaki scheme on integer matrix multiplication unit (IMMU) for high-performance computing applications.
By using IMMU, especially INT8-input-INT32-accumulation IMMU, we can reduce the memory consumption and the number of operations while keeping the same accuracy as the FMMU.
In addition to these advantages, the theoretical peak throughput of IMMUs is typically higher than that of FMMUs, allowing for further speedups.
We implement the Ozaki scheme on NVIDIA integer Tensor Cores and evaluate the accuracy, throughput, and power consumption.
Our implementation outperforms cuBLAS DGEMM and existing Ozaki scheme implementation on FP16 Tensor Cores on consumer GPUs while keeping the accuracy and consuming lower power.
Moreover, we apply the Ozaki scheme on integer Tensor Cores to quantum circuit simulation and achieve up to a $4.33$ times throughput improvement compared to cuBLAS ZGEMM implementation while managing accuracy.
Overall, our results demonstrate the advantages of utilizing IMMU for HPC applications and showcase the potential of the Ozaki scheme on integer Tensor Cores for accelerating scientific computing workloads.

\section{Discussion}
\textit{Is the Ozaki scheme a superior alternative to DGEMM in all applications?}
In other words, \textit{does it consistently provide better or equivalent accuracy compared to DGEMM while achieving higher throughput?}
Our response is no; its suitability depends on both the problem at hand and the performance of the hardware being used.
The factors influencing numerical errors in floating-point operations and the Ozaki scheme are distinct, and there exists a trade-off between accuracy and throughput within the Ozaki scheme.
The Ozaki scheme can be faster than DGEMM by floating-point operations depending on a problem, specifically when the distribution of the exponent is relatively small, as we mentioned in the \refname{sec:ozaki-disadvantage} section of the Ozaki scheme explanation.
Consequently, it is essential to conduct further investigations into the exponent distribution that arises within a given application to determine the appropriateness of applying the Ozaki scheme.
Additionally, the extent to which the Ozaki scheme is faster is contingent on the hardware's performance capabilities.

To maintain the balance between accuracy and throughput in our evaluation of the Ozaki scheme using Int8 Tensor Cores within quantum circuit simulations, we employ a method to determine the number of splits by scanning all elements of the matrices.
This method relies on an average mantissa loss length as a selection threshold.
Then, \textit{is this criterion truly optimal for selecting the number of splits?}
Our answer is a resounding no; it does not yield the optimal number of splits when it comes to the exact DGEMM emulation.
This is because of the fundamental differences in error sources between floating-point operations and the Ozaki scheme.
In matrix multiplication using floating-point operations, the accumulated rounding error increases as the accumulation length grows.
In contrast, with the Ozaki scheme, the rounding error remains stable even with an increase in the accumulation length of matrix multiplication.
Instead, it is the insufficient number of splits that adversely affects numerical accuracy.
Therefore, the accumulation length should be one of the key factors in determining the optimal number of splits in the Ozaki scheme to emulate DGEMM by floating-point operations exactly, which we did not consider.
Another factor we need to consider if we emulate DGEMM by floating-point operations exactly is the treatment of values whose absolute values are small.
In the method we have used to determine the number of splits, the matrix elements whose absolute value is relatively small are underflowed and treated as zeros when splitting the matrix.
This is one of the weaknesses of the fixed-point representation over floating-point.
Whether this results in significant accuracy degradation depends on the interplay between these elements and those of the matrix to which they are multiplied.
To illustrate this point, consider two scenarios:
\begin{enumerate}
    \item In an inner product calculation between vectors $a=\begin{bmatrix}1 & 1\end{bmatrix}$ and $b=\begin{bmatrix}1 & 10^{-20}\end{bmatrix}$ in double-precision, the value $10^{-20}$ in vector $b$ can reasonably be regarded as zero. This is because the result remains virtually unchanged even if we treat it as zero.
    \item Conversely, in the case of vectors $a=\begin{bmatrix}1 & 10^{20}\end{bmatrix}$ and $b=\begin{bmatrix}1 & 10^{-20}\end{bmatrix}$ in double-precision, the $10^{-20}$ value in vector $b$ cannot be treated as zero, as it significantly affects the result.
\end{enumerate}
The problem is that conducting an exhaustive examination of all matrix elements and those involved in multiplication demands considerably more computational resources than standard matrix multiplication itself.
This approach would be akin to putting the cart before the horse.
We opt for the simplicity and low computational overhead of our method for selecting the number of splits.
Exploring and implementing a more efficient algorithm for determining the optimal number of splits stands as a viable pathway to enhancing throughput and harnessing the full potential of the Ozaki scheme in various applications.

\begin{acks}
We would like to thank Hidetaka Manabe for helpful discussions and comments.
This work was supported by JSPS KAKENHI Grant Number JP20H04195, JP20K20624, JP21H03447, JP22H03598. This work is supported by JST CREST Grant Number JPMJCR2112. This work is supported by "Joint Usage/Research Center for Interdisciplinary Large-scale Information Infrastructures" in Japan (Project ID: jh230053-NAH, jh230009-NAHI).
\end{acks}

\bibliographystyle{plain} 
\bibliography{references}

\end{document}